\documentclass[12pt]{article}
\usepackage{amsmath,amssymb,epsfig}
\makeatletter \@addtoreset{equation}{section} \makeatother
\renewcommand{\theequation}{\thesection.\arabic{equation}}
\newcommand{\ba}{\begin{array}}
\newcommand{\ea}{\end{array}}
\newcommand{\beq}{\begin{equation}}
\newcommand{\eeq}{\end{equation}}
\newcommand{\bea}{\begin{eqnarray}}
\newcommand{\eea}{\end{eqnarray}}

\def\bce{\begin{center}}
\def\ece{\end{center}}

\def\nonu{\nonumber}

\def\be{\beta}

\newcommand{\tr}{\mbox{Tr}}

\def\eps6{{\displaystyle \mathop{\epsilon}^{6}}{}}

\def\nab6{{\displaystyle \mathop{\nabla}^{6}}{}}

\def\0{{\sst{(0)}}}
\def\1{{\sst{(1)}}}
\def\2{{\sst{(2)}}}
\def\3{{\sst{(3)}}}
\def\4{{\sst{(4)}}}
\def\5{{\sst{(5)}}}
\def\6{{\sst{(6)}}}
\def\7{{\sst{(7)}}}
\def\8{{\sst{(8)}}}

\def\ba{\begin{array}}
\def\ea{\end{array}}
\def\beq{\begin{equation}}
\def\eeq{\end{equation}}
\def\be{\begin{equation}}
\def\ee{\end{equation}}

\def\tr{\mathop{\rm tr}}

\def\eps{\epsilon}

\def\ba{\begin{array}}
\def\ea{\end{array}}
\def\beq{\begin{equation}}
\def\eeq{\end{equation}}
\def\be{\begin{equation}}
\def\ee{\end{equation}}

\def\tr{\mathop{\rm tr}}

\def\eps{\epsilon}

\newcommand{\bean}{\begin{eqnarray*}}
\newcommand{\eean}{\end{eqnarray*}}

\begin{document}
\thispagestyle{empty} \addtocounter{page}{-1}
\begin{flushright}
{\tt hep-th/0701145}\\
\end{flushright}

\vspace*{1.3cm}

\centerline{ \Large \bf Meta-Stable Brane Configuration  with Orientifold 6 Plane}
\vspace*{1.5cm}
\centerline{{\bf Changhyun Ahn} 
} 
\vspace*{1.0cm} 
\centerline{\it 
Department of Physics, Kyungpook National University, Taegu
702-701, Korea} 
\vspace*{0.8cm} 
\centerline{\tt ahn@knu.ac.kr} 
\vskip2cm

\centerline{\bf Abstract}
\vspace*{0.5cm}

We present the intersecting brane configuration of type IIA string
theory corresponding to 
the meta-stable nonsupersymmetric vacua in four dimensional ${\cal
  N}=1$
supersymmetric $SU(N_c)$ gauge theory with   
a symmetric flavor, a conjugate symmetric flavor and 
fundamental flavors.   By studying the previously known supersymmetric 
M5-brane curve, the M-theory lift for this type IIA brane configuration,
which consists of NS5-branes, D4-branes, D6-branes and an orientifold
6-plane, is analyzed.  

\baselineskip=18pt
\newpage
\renewcommand{\theequation}
{\arabic{section}\mbox{.}\arabic{equation}}

\section{Introduction}

The dynamical supersymmetry breaking \cite{ISS} occurs in ${\cal N}=1$ SQCD with
massive fundamental flavor where the masses are much smaller than the
dynamical scale of the gauge sector. The supersymmetry is broken by
the rank condition: not all the F-term equations can be
satisfied. Contrary to the electric theory which has a vanishing
superpotential, 
the dual magnetic theory has a
superpotential term interacting between the magnetic mesons and the magnetic
quarks. See, for example, \cite{GK}. 
In this construction of ISS model \cite{ISS}, 
both the mass for the flavors and  the Seiberg duality
play an important role.
By deforming the electric theory through the mass term for quarks, 
one can add a superpotential. This can be interpreted as a linear term
in magnetic mesons in the magnetic theory. Then, the F-term
conditions in the dual magnetic theory
can be obtained from these two contributions, linear term in
magnetic mesons plus the Yukawa coupling term above.
It turns out that the F-terms of the mesons cannot all vanish due to
the fact that the ranks are different.    
On the other hand, the definition of 
meta-stable states is meaningful only when they are long-lived
parametrically. For the particular range of the number of flavors
where the Seiberg dual should be applied,
one can make the meta-stable vacuum long-lived arbitrarily.  
Therefore, it is necessary to consider both the mass for the flavors and
Seiberg dual magnetic theory in order to find out new meta-stable
supersymmetry breaking vacua.

In the type IIA brane configuration, the quark masses correspond to 
the relative displacement of the D6-branes (0123789) and D4-branes
(01236) along the 45 directions geometrically. The eigenvalues of mass matrix
correspond to the locations of D6-branes in 45 directions. See, for
example,  
\cite{GK}.
On the other hand, the Seiberg duality in the classical brane picture 
can be achieved by exchanging the locations of the NS5-brane (012345) and 
NS5'-brane (012389) along $x^6$ direction each other \cite{GK}. 
The geometric misalignment of D4-branes connecting both NS5'-brane  and
D6-branes in the magnetic brane configuration 
can be interpreted as a nontrivial F-term conditions in the
gauge theory side \cite{OO,FGU,BGHSS,Ahn06}. 
Then the F-term can be partially cancelled by both recombination of 
flavor D4-branes (connecting to NS5'-brane) with the color 
D4-branes (connecting to
NS5-brane) and then movement of those into the 45 directions.
This phenomenon in brane configuration 
corresponds to the fact that some entries in the dual
quarks
acquire nonzero vacuum expectation values to minimize the F-term in
the gauge theory side. 
Since there  are no D4-branes between NS5-brane and NS5'-brane, 
the magnetic gauge group is completely Higgsed. 
Moreover, the remaining flavor D4-branes connecting to 
NS5'-brane can move along 89 directions freely and independently since
D6-branes and NS5'-brane are parallel
and 
this geometric freedom corresponds to the
classical pseudomoduli space of nonsupersymmetric vacua of the gauge theory. 

In this paper, we study ${\cal N}=1$ $SU(N_c)$ gauge theory with 
a symmetric flavor $S$, a conjugate symmetric flavor $\widetilde{S}$
and 
$N_f$ fundamental
flavors $Q$ and $\widetilde{Q}$ 
\footnote{The type IIA brane configuration describing the
nonsupersymmetric meta-stable minimum in this theory is proposed in 
\cite{FGU}.} in the context of dynamical supersymmetric breaking vacua.
The corresponding supersymmetric brane configuration in type IIA
string theory was given by Landsteiner, Lopez
and Lowe in \cite{LLL} sometime ago.
In the gauge theory analysis \cite{ILS} alone, the turning on the superpotential 
$W =  \tr (S \widetilde{S})^2$ was necessary to truncate the chiral
ring and the dual description was given. 
From the brane
analysis \cite{LLL} where, in general, two other quartic terms in the
superpotential are
present, 
this quartic superpotential can be obtained from
the corresponding ${\cal N}=2$ theory by integrating the massive
adjoint field out \cite{GK}. 

The point is that for the limit of
infinite mass for the adjoint field,
the above quartic superpotential vanishes, because the coefficient
function appears as an inverse of mass, as in the electric theory of
${\cal N}=1$ SQCD with flavors we mentioned before. 
Now we can add the mass
term for quarks in the fundamental representation of the gauge group
while keeping the infinite mass limit. 
Then we turn to the
dual magnetic  gauge theories by two standard brane motions \cite{GK} with
appropriate linking number counting. It
turns out that the dual
magnetic theory giving rise to the meta-stable vacua 
is described by ${\cal N}=1$
$SU(2N_f-N_c)$ gauge theory with dual matter contents and the
superpotential consists of  linear term in
magnetic meson plus the coupling terms between the magnetic meson,
dual quarks, dual symmetric tensor flavor, and dual conjugate
symmetric tensor flavor (\ref{correctsuperpotential}). 
Some of the F-term conditions cannot be satisfied by rank condition.
The final nonsupersymmetric minimal energy brane configuration for
this theory is shown in Figure 3.  

In section 2, we review the type IIA brane configuration corresponding
to the electric theory based on the ${\cal N}=1$ $SU(N_c)$ gauge theory
with above matter contents. 
In section 3, we construct the Seiberg dual magnetic theory which is 
${\cal N}=1$ $SU(3N_f-N_c+4)$ gauge theory with corresponding dual
matters as well as various gauge singlets, by brane motion and linking
number counting.
In section 4, we consider the infinite mass limit for the adjoint field of
${\cal N}=2$ theory in order to obtain nonsupersymmetric meta-stable
minimum where the gauge group is given by $SU(2N_f-N_c)$ with matter
contents
and present 
the corresponding intersecting brane configuration of type IIA string
theory.
In section 5, we describe M-theory lift of the supersymmetry breaking 
type IIA brane configuration we have considered in section 4, by
following the work of \cite{BGHSS,Ahn06-1}.
Finally, in section 6, we give the summary of this paper and 
make some comments for the future directions. 

For the relevant works on the meta-stable vacua in various and different contexts, 
we refer to some
partial lists appeared in \cite{FU}-\cite{AGM1}. 

\section{The ${\cal N}=1$ supersymmetric electric brane configuration}

The Seiberg-Witten curve of ${\cal N}=2$ $SU(N_c)$ gauge theory with 
a symmetric flavor $S$, a conjugate symmetric flavor $\widetilde{S}$ and 
$N_f$ fundamental flavors $Q, \widetilde{Q}$ was found in
\cite{LLL} \footnote{For an asymptotic free region \cite{LLL}, we should have
  $N_c >N_f+2$. }.
The corresponding type IIA 
brane configuration \cite{LL} consists of three NS5-branes (012345)
which have different $x^6$ values, $N_c$
D4-branes (01236)
suspended between them, $2N_f$ D6-branes (0123789) and an
orientifold 6 plane (0123789) of positive Ramond charge where a middle
NS5-brane is located \footnote{The orientifold action identifies the two
factors of gauge group $SU(N_c) \times SU(N_c)$ and projects the
bifundamental representation onto the symmetric representation
\cite{LL,LLL}. The brane configuration of ${\cal N}=1$ supersymmetric
gauge theory with $SU(N_L) \times SU(N_R)$ and matter in the
bifundamental and fundamental representations was found in
\cite{BH,GP}. See also the relevant work in \cite{ENR}.}. 
Let us introduce two complex coordinates
\bea
v \equiv x^4 + i x^5, \qquad w \equiv x^8 + i x^9.
\nonu
\eea
We'll introduce these more precise way later when we consider 11-dimensional theory.
According to ${\bf Z}_2$ symmetry of orientifold
6-plane(O6-plane) \footnote{For the negative Ramond
charge of orientifold 6-plane with same brane configuration \cite{LL}, 
the matter contents in the gauge theory side are changed into 
an antisymmetric flavor $A$ and a conjugate antisymmetric flavor
$\widetilde{A}$ as well as fundamentals $Q$ and $\widetilde{Q}$. 
We'll concentrate on the case with positive O6-plane
and the other case with negative O6-plane will be similar to what we
present here and can be
done without any difficulty by recognizing the dependence on the negative charge in
various places. } 
sitting at $v=0$ and $x^6=0$,
the coordinates $(x^4,x^5, x^6)$ transform as $-(x^4, x^5, x^6)$. 
The left NS5-brane with positive (negative) $v$ is mirror to the right NS5-brane
with negative (positive) $v$ and the middle NS5-brane with positive $v$
is mirror to the middle NS5-brane with negative $v$. Every D4-brane
and D6-brane which do not pass through $v=0$ should also have its
mirror image, according to this rule. 
The distance along $v$ direction of D4 branes is related
to the mass of two index tensor matter ($S \widetilde{S}$) and we put
all the D4-branes at $v=0$. Of course, the D4-branes with positive $x^6$ are
mirror to those with negative $x^6$.  Then the distance along $v$ direction of
D6-branes provides the mass for the fundamental matter ($Q
\widetilde{Q}$). 
For the equal
masses, the $v$ coordinate of $N_f$ D6-branes with nonzero $x^6$ has
only one fixed value
and its $N_f$  mirrors appear with opposite values of $v$ and $x^6$. 
See Figure 1 for clear view on the brane configuration.

By rotating the left and right NS5-branes from $v$ direction toward
$\pm w$ 
direction (equivalently introducing the mass for adjoint field of
${\cal N}=2$ theory and integrating out this massive adjoint field \cite{GK})
respectively, 
one obtains 
${\cal N}=1$ theory.  
Among three NS5-branes, the middle NS5-brane is stuck on an 
O6-plane as in Figure 1 and the outer two NS5-branes can be rotated in a
${\bf Z}_2$ symmetric manner due to the presence of orientifold 6
plane. 
That is, if the left NS5-brane rotates by an angle $\theta$ in
$(v,w)$ plane, denoted by  $NS5_{\theta}$-brane \cite{GK}, the mirror image of
this NS5-brane, the right NS5-brane, is rotated by an angle $-\theta$
in the same plane, denoted by 
$NS5_{-\theta}$-brane.  For more details, see the Figure 1.
At the moment, the angle $\theta$ is greater than 0 and
less than $\pi/2$. We will come to the case for $\theta=\pi/2$ later section.
We also rotate the $N_f$ D6-branes and make them be parallel to
$NS5_{\theta}$ and denote them as $D6_{\theta}$(the notation is not
good because the
angle between the unrotated D6-branes and $D6_{\theta}$-branes is
equal to $\pi/2-\theta$, not $\theta$) and its mirrors $N_f$ D6-branes appear
as $D6_{-\theta}$. There is no coupling between the adjoint field and
the quarks since the rotated $D6_{\theta}$-branes are parallel to the rotated
$NS5_{\theta}$-brane \cite{BH,GK}.
For this brane setup, the classical superpotential is
given by a quartic term for tensor matter and mass term for quarks as follows 
\footnote{If there exist $k$ coincident $NS5_{\theta}$-branes (and
  their mirrors also), then the superpotential with massless quarks
  takes the form $W=\tr (S \widetilde{S})^{k+1}$ and the corresponding
  field theory analysis on the magnetic dual was given in
  \cite{ILS}. One can understand the quartic superpotential by writing
  the full superpotential as $W=\mu \Phi^2 + S \Phi \widetilde{S}$ and
  integrating the adjoint field $\Phi$ out.  
\label{kgeneral}}
\bea
W = \frac{1}{\mu} \tr (S \widetilde{S})^2 + \tr m Q \widetilde{Q}
\nonu 
\eea 
where $S$ and
$\widetilde{S}$ represent  a flavor of symmetric and a conjugate
symmetric 
tensor representations 
of $SU(N_c)$ and the adjoint mass $\mu$ is given by the rotation angle $\theta$
through \cite{GK} 
\bea
\mu \equiv \tan \theta.
\nonu
\eea

The field theory analysis \cite{ILS} \footnote{When the baryonic
  operator $B_n = S^n Q^{N_c-n} Q^{N_c-n}$ gets an expectation value,
  then the initial gauge group $SU(N_c)$ is broken to $SO(n)$ with a
  symmetric tensor $\widetilde{S}$ and $2N_f$ vectors and the
  superpotential will be $\tr \widetilde{S}^2$ \cite{ILS}. Similarly, 
 when the baryonic
  operator $B_n = A^n Q^{N_c-2n}$ gets an expectation value with
  negative O6-plane charge,
  then the initial gauge group $SU(N_c)$ is broken to $Sp(n)$ with a conjugate
  antisymmetric tensor $\widetilde{A}$ and $N_f$ flavors and the
  superpotential will be $\tr \widetilde{A}^2$.
\label{baryon}}
corresponding to this superpotential which
was necessary to truncate the chiral ring (with
massless quarks) provides the dual description. 
Note that there exists an extra global $U(1)_S$ symmetry acting on $S$ and
$\widetilde{S}$ only
and is not realized geometrically in the brane configuration \cite{LLL}. 
The masses of quarks are given by the location of D6-branes in $v$
direction (Note that all the D4-branes are located at $v=0$) as we
mentioned before. 
If we leave the D6-branes parallel to the O6-plane, there will be more
terms in the superpotential \cite{LLL} \footnote{According to the result
  of \cite{CSST}, the dual magnetic gauge group from brane
  configuration
is given by $SU(2N_f-N_c+4)$ and
the superpotential has more terms with the same matter contents as those
in electric theory. Of course, these extra two terms in the
superpotential can be interpreted as mesonic perturbation in the
magnetic theory \cite{CSST}.  }.
After this rotation, the three NS5-branes intersect at a point in (4589) directions
implying that the mass for symmetric flavor or conjugate symmetric
flavor($S \widetilde{S}$) vanishes.
For the infinite limit of $\mu$, the superpotential becomes 
$\tr m Q \widetilde{Q}$ which is the same superpotential for the
${\cal N}=1$ SQCD with massive flavors. 
For the more general brane configurations with nonzero $S
\widetilde{S}$, 
we refer to \cite{LLL,GP}. 

Now one summarizes
the supersymmetric electric brane configuration with their
worldvolumes in type IIA string theory as follows.

$\bullet$
Middle NS5-brane with worldvolume (012345) with $w=0=x^6$.

$\bullet$
Left $NS5_{\theta}$-brane with worldvolume by both (0123) and two spatial dimensions
in $(v,w)$ plane and with negative $x^6$.

$\bullet$
Right $NS5_{-\theta}$-brane  with worldvolume by both (0123) and two spatial dimensions
in $(v,w)$ plane and with positive $x^6$.

$\bullet$ Left $N_f$
 $D6_{\theta}$-branes  with worldvolume by both (01237) and two spatial dimensions
in $(v,w)$ plane and with negative $x^6$. Before the rotation, the
 distance from D4-branes in the $v$ direction is nonzero.

$\bullet$ Right $N_f$
$D6_{-\theta}$-branes  with worldvolume by both (01237) and two space dimensions
in $(v,w)$ plane and with positive $x^6$.

$\bullet$ O6-plane  with worldvolume (0123789) with $v=0=x^6$.

$\bullet$ $N_c$ D4-branes  with worldvolume (01236) with $v=0=w$.

We draw the type IIA electric brane configuration in Figure 1 which was basically given in
\cite{LLL,GK} already but the only difference is to put D6-branes in the
nonzero $v$ direction in order to obtain nonzero masses for the quarks
which are necessary to obtain the meta-stable vacua as we observed in
the previous section
\footnote{One of the variants of this brane configuration (no middle
  NS5-brane) 
provides ${\cal
    N}=1$ symplectic or orthogonal gauge groups with $N_f$ flavors depending
  on the charges of O6-plane with a quartic superpotential (after
  integrating out the massive adjoint field) for the
  quarks: 
Left $NS5_{\theta}$-brane, Right
  $NS5_{-\theta}$-brane,
$2N_f$ D6-branes which are not parallel to $NS5_{\pm \theta}$-branes,  
O6-plane and  $N_c$ D4-branes \cite{CSST,AOT,GK}. It is not clear how
the meta-stable brane construction from the magnetic dual description
of this theory with infinite mass for the adjoint field
arises. Without D6-branes, this theory has a superpotential
$W=\frac{1}{\mu} \tr (S
\widetilde{S})^2$, as usual.
    }. Also one can imagine the brane configuration for $\theta=\pi/2$
from Figure 1.  

\begin{figure}[ht]
   \epsfxsize=6in 
\centerline{\epsffile{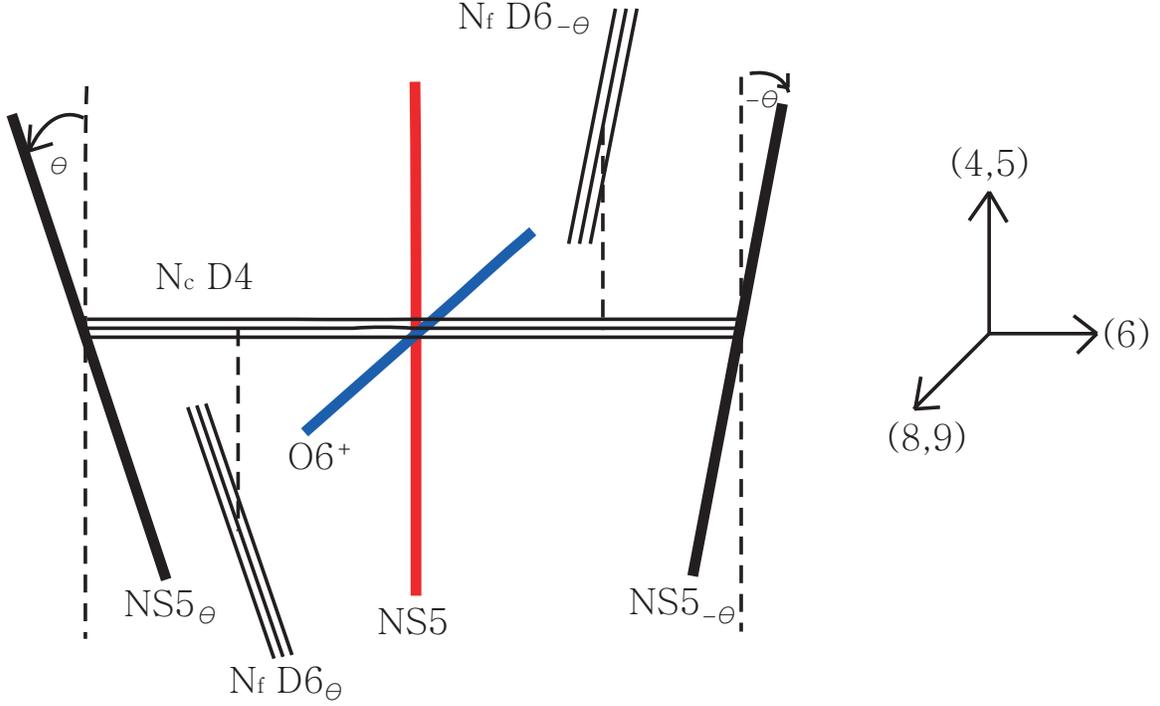}}
   \caption[FIG. \arabic{figure}.]{ 
The ${\cal N}=1$ 
supersymmetric electric brane configuration for the $SU(N_c)$ gauge
theory  with 
a symmetric flavor $S$, a conjugate symmetric flavor $\widetilde{S}$
(which are massless) that are strings stretching between D4-branes
located at the left hand side of O6-plane and those at the right hand
side of O6-plane  and 
$N_f$ fundamental massive flavors $Q, \widetilde{Q}$ that are strings
stretching between D6-branes and D4-branes. 
The origin of
the coordinates $(x^6, v, w)$ is located at the intersection of
NS5-brane and O6-plane. }
\label{fig1}
\end{figure}

\section{The ${\cal N}=1$ supersymmetric magnetic brane configuration}

As explained in the introduction, 
the Seiberg duality is crucial for the construction of meta-stable vacua.
In order to obtain the magnetic dual theory in the context of brane
configuration, 
we move the D6 and
NS5-branes through each other \cite{GK} and use the linking numbers \cite{HW} for the
computation of D4-branes which are created during this process.
In other words, the magnetic theory in the brane picture is obtained 
by interchanging D6-branes and NS5-branes with their mirror images 
while preserving the linking number.
The linking number \cite{HW} for the D6-branes is given by \footnote{In
  \cite{GK}, the linking number has opposite sign but the overall sign
  is not important when we use the conservation of linking number
  between before the brane motion and after the brane motion.} 
\bea
L_6 = \frac{1}{2}(n_{5L}-n_{5R}) +n_{4R}-n_{4L}
\label{l6}
\eea
where $n_{5L,R}$ are the NS5-branes to the left or right of the
D6-branes and  $n_{4L,R}$ are the D4-branes to the left or right of the
D6-branes.
After we move the left $D6_{\theta}$-branes to the right all the way(and their
mirrors, right $D6_{-\theta}$-branes to the left) past the NS5-brane and
$NS5_{-\theta}$-brane, 
the linking number $L_6$ of a $D6_{\theta}$-brane
becomes $L_6=1-n_{4L}$, because $n_{5L}=2$ and $n_{5R}=0=n_{4R}$, 
which should be equal to the original 
$L_6=-1$  because $n_{5L}=0=n_{4L}=n_{4R}$ 
and $n_{5R}=2$ before the brane motion, 
according to the conservation of linking number. 
Then the $n_{4L}$ becomes 2 and we must add $2N_f$ D4-branes to the
left side of all $N_f$ $D6_{\theta}$-branes (and their mirrors). See
Figure 2 for the creation of these D4-branes. 

Next, we move the left $NS5_{\theta}$-brane to the right all the way past
O6-plane
(and its mirror, right $NS5_{-\theta}$-brane to the left), the linking number
of $NS5_{\theta}$-brane, 
$L_5 =\frac{N_f}{2}+\frac{1}{2} (4)-N+ 2 N_f$
using the general formula \footnote{One can also add the contribution
  from O6-plane explicitly as in \cite{EGKRS}. In our convention,
  $n_{6L,R}$ includes the linking number from O6-plane.  }
\bea
L_5 = \frac{1}{2}(n_{6L}-n_{6R}) +n_{4R}-n_{4L}
\label{l5}
\eea
where $n_{6L,R}$ are the D6-branes to the left or right of the
NS5-branes, $n_{4L,R}$ are the D4-branes to the left or right of the
NS5-branes and the linking number for O6-plane is given by 4 D6-branes. 
Originally it was $L_5=-\frac{N_f}{2}-\frac{1}{2}(4)
+N_c$ from Figure 1 before the brane motion.
This leads to the fact that the number of D4-branes along the
$x^6$ direction becomes
\bea
N=3N_f-N_c+4
\label{general}
\eea 
for the dual magnetic gauge group
\footnote{ For magnetic IR free region, we should have $N_f > N-2$ and
this implies that $\frac{N_c}{3}-\frac{4}{3} < N_f < \frac{N_c}{2} -1$.}. 
This result \footnote{It is natural to ask whether this analysis can
  be generalized to the arbitrary superpotential given in the footnote 
\ref{kgeneral}. It is straightforward to take the Seiberg dual
successively and it turns out one arrives at the dual magnetic gauge group
$SU((2k+1)N_f -N_c +4k)$ where $2 k N_f$ term comes from $k$ coincident
$NS5_{\pm \theta}$-branes, $N_f$ term comes from a middle NS5-brane
and the presence of $4k$ in the dual gauge group is 
due to the O6-plane \cite{GK}. } 
agrees with the field theory analysis in \cite{ILS}.
It is
easy to check that the linking numbers for the middle NS5-brane are
consistent
with this dual configuration.
The dual brane configuration \footnote{The linking number counting
  without O6-plane was given in \cite{BH} and similar counting without
a middle NS5-brane was given in \cite{CSST}. Also other example of
linking number counting appears in \cite{LLL1}. } 
is shown in Figure 2.
In particular, $2N_f$ D4-branes connecting between
$NS5_{-\theta}$-brane and $D6_{-\theta}$-branes are the singlets
corresponding to the mesons 
\bea
M_0 \equiv Q\widetilde{Q} \qquad 
\mbox{and} \qquad M_1\equiv Q \widetilde{S} S \widetilde{Q}.
\label{mesons}
\eea
The former occurs when $D6_{\theta}$-branes cross $NS5_{-\theta}$-brane while 
the latter  occurs when $D6_{\theta}$-branes cross a middle NS5-brane.
They are gauge singlets under $SU(N)$ and are fundamentals $({\bf N_f}, {\bf N_f})$ 
under the global flavor symmetry $SU(N_f)_L \times SU(N_f)_R$.
From the ${\cal N}=1$ electric/magnetic duality we expect to get 
the dual gauge group
$SU(3N_f-N_c+4)$  
 with 
a symmetric flavor $s$, a conjugate symmetric flavor $\widetilde{s}$ and 
$N_f$ fundamental flavors $q, \widetilde{q}$ as well as various gauge singlets.
By adding a linear term in $M_0$, the superpotential term \cite{ILS}
becomes
\bea
W_{dual} = (s \widetilde{s})^2 + M_0 q  \widetilde{s} s
\widetilde{q} + M_1 q \widetilde{q}+ P_0 q \widetilde{s} q +
\widetilde{P}_0 \widetilde{q} s \widetilde{q} + m M_0
\label{dual}
\eea
where two other singlets are given by 
$P_0 \equiv Q \widetilde{S} Q$ and $\widetilde{P}_0 \equiv \widetilde{Q} S
\widetilde{Q}$ which are symmetric in their flavor indices
\footnote{When D6-branes are unrotated \cite{CSST} and not parallel to
rotated NS5-branes, then in the superpotential there exist extra terms
$(Q\widetilde{Q})^2 + Q \widetilde{S} S \widetilde{Q}$ in the electric
theory and these will appear as $M_0^2+M_1$ in the magnetic theory
from the point of brane configuration, as pointed out in \cite{CSST}.
\label{more}}. 
This theory has the $SU(N_f)_L \times SU(N_f)_R \times U(1)_S \times
U(1)_B 
\times U(1)_R$
global flavor symmetry.
 The explicit matter field transformations under these symmetries 
are given in \cite{ILS}.
In the dual theory \footnote{It was found in \cite{ILS} that with the
  deformation for the mass of $N_f$-th quark, the vacuum giving the
  expectation values $<q_{N_f}>, <\widetilde{q}_{N_f}>, <s>$ and
  $<\widetilde{s}>$
breaks the magnetic gauge group 
$SU(3N_f-N_c+4)$ into $SU(3N_f-N_c+1)$ with $(N_f-1)$ remaining light flavors.
The $M_0$ and $M_1$ equations of motion imply this vacuum of the
theory. Of course, for the more general superpotential (more than
quartic superpotential), this kind of
deformation is applicable.}, 
the operator $B_n$ defined in footnote
\ref{baryon} is mapped to $b_n=s^{(2N_f+4)-n} q^{N_f+n-N_c}
q^{N_f+n-N_c}$
and when this operator gets an expectation value, the dual $SU(N)$
gauge theory is broken to $SO(\widetilde{n}=2N_f-n+4)$ and 
the low energy has $2N_f$ matter fields and a symmetric tensor \cite{ILS}. 
If we turn on the operator $\widetilde{B}_n = \widetilde{S}^n 
\widetilde{Q}^{N_c-n} \widetilde{Q}^{N_c-n}$ getting the expectation
values further in the electric theory, then the initial gauge group is
broken to $SO(n)$ with $2N_f$ vectors and the dual description on this
operator can be done similarly.  

\begin{figure}[ht]
   \epsfxsize=6in 
\centerline{\epsffile{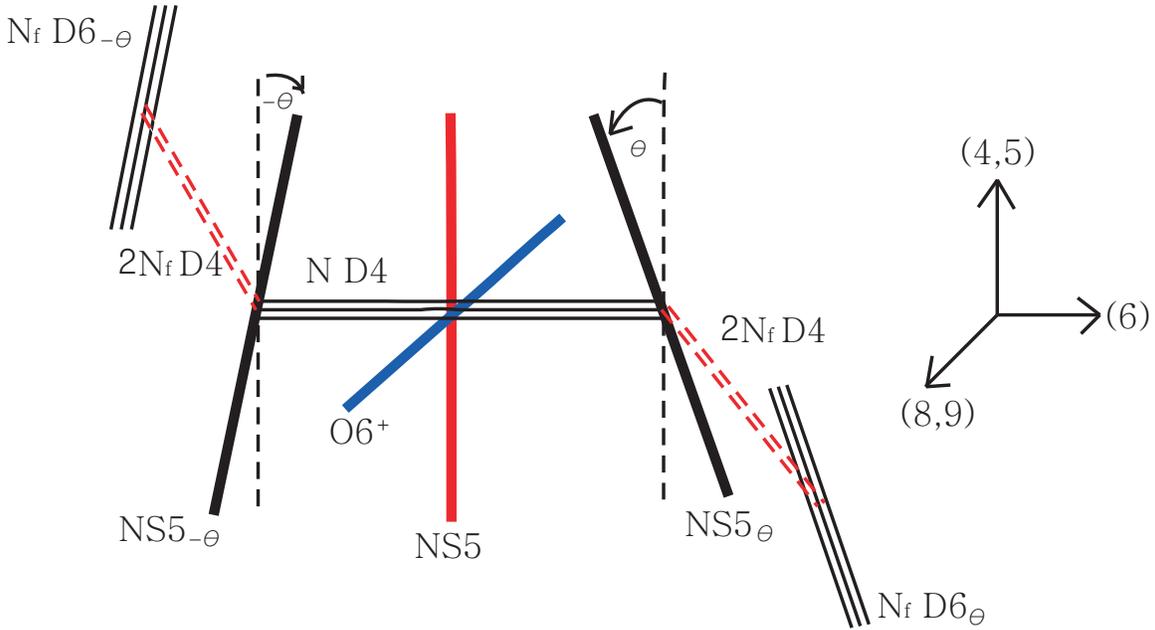}}
   \caption[FIG. \arabic{figure}.]{ 
The ${\cal N}=1$ supersymmetric magnetic 
brane configuration for the $SU(N=3N_f-N_c+4)$ gauge
theory  with 
a symmetric flavor $s$, a conjugate symmetric flavor $\widetilde{s}$ and 
$N_f$ fundamental flavors $q, \widetilde{q}$.
 The $2N_f$ D4-branes connecting between
$NS5_{-\theta}$-brane and $D6_{-\theta}$-branes are the dual gauge singlets
corresponding to the mesons $M_0$ and $M_1$ (\ref{mesons}). 
As will be explained in section
\ref{four},
in the $\mu \rightarrow \infty$ limit (or $\theta \rightarrow \pi/2$ limit),
only $N_f$ D4-branes connecting NS5'-brane and D6-branes appear (and
their mirrors). In the superpotential, it contains either $M_0$ or
$M_1$-dependent term depending on which way the brane motion occurs.}
\label{fig2}
\end{figure}

\section{Nonsupersymmetric meta-stable brane
  configuration 
\label{four}}

We have seen that
there are $2N_f$ D4-branes connecting between
$NS5_{-\theta}$-brane and $D6_{-\theta}$-branes (and their mirrors) for
finite $\mu$ in
Figure 2.
In order to describe $\mu=\infty$ limit case in this section,
one needs to see each step from the electric theory to magnetic theory
more closely. 
For fixed left and right NS5'-branes, the D6-branes can be located 
either at the parallel position with NS5'-branes or at the nonparallel
position
with them in the electric theory. 
We are interested in the case where they are perpendicular
or parallel to each other because only these two cases provide the
situation
where the above $2N_f$ D4-branes creation can be decomposed into half
$N_f$ D4-branes and half $N_f$ D4-branes independently.
Let us describe two different cases separately as follows.

{\bf 1. When the D6-branes are perpendicular to a middle NS5-brane in
  the electric theory}
 
So far, the rotation angle is less than $\pi/2$. In this section,
we consider the case of $\theta=\pi/2$. Then the left
$NS5_{\theta}$-brane in the electric theory 
is parallel to the right $NS5_{-\theta}$-brane
and both $D6_{\pm \theta}$-branes 
become unrotated D6-branes which are parallel to the above $NS5'
\equiv NS5_{\pm \pi/2}$ brane.
During the process of previous section when $\theta=\pi/2$, 
the creation of D4-branes is
a little bit changed because when a D6-brane crosses over
NS5'-brane and when NS5'-brane crosses O6-plane or D6-brane, 
there is no creation of D4-branes.  
In other words, a D4-brane is created whenever a NS5-brane crosses
a D6-brane as long as the NS5-brane and a D6-brane are not parallel.

After we move the left D6-branes to the right all the way (and their
mirrors, right D6-branes to the left) past the middle NS5-brane and
the right NS5'-brane, 
the linking number $L_6$ of a single D6-brane from (\ref{l6})
becomes $L_6=\frac{1}{2}-n_{4L}$, since $n_{5L}=1$ from a middle NS5-brane, 
which should be equal to the original linking number 
$L_6=-\frac{1}{2}$ because $n_{5R}=1$ from a middle NS5-brane. 
Then we should add $N_f$ D4-branes, corresponding to the meson $M_1$, to the left 
side of all the right $N_f$ D6-branes (and their mirrors) in Figure 2. 
Note that the difference, compared with previous case where $0<\theta
< \pi/2$, 
appears here when a D6-brane cross the right NS5'-brane.
Due to the parallelness of these, there is no creation of D4-branes.

Next, we move the left NS5'-brane to the right all the way past
O6-plane
(and its mirror, right NS5'-brane to the left), and then the linking number
of NS5'-brane from (\ref{l5}), $L_5 =-N +  N_f$ from $n_{4L}=N$ and $n_{4R}=N_f$.
Originally, it was $L_5=N_c$ because $n_{4R}=N_c$ in Figure 1.
This implies that $N$ becomes 
\bea
N=N_f-N_c.
\label{caseone}
\eea
There is no D4-brane creation when NS5'-brane crosses an O6-plane
because they are parallel to each other. This is a new feature,
compared with the case of finite $\mu$. 

Finally, we are
left with Figure 2 with $\theta=\pi/2$ 
except that the number of D4-branes connecting
between D6-branes and NS5'-brane is $N_f$ not $2N_f$ and the gauge
group is $SU(N=N_f-N_c)$.
These $N_f$ D4-branes correspond to the singlet $M_1$ (\ref{mesons}).
In this case,
the superpotential is given by
$
W_{dual} = \tr M_1 q \widetilde{q} + \tr m M_0
$
where the first term comes from the superpotential with $M_1$ for
finite $\mu$ case in previous section (\ref{dual}) and the second term comes from
the mass term for the quarks in the context of magnetic theory. 
This magnetic theory doesn't produce the meta-stable vacua because all
the F-term equations are satisfied.

{\bf 2. When the D6-branes are parallel to a middle NS5-brane or 
perpendicular to NS5'-branes in the electric theory}

Let us first rotate D6-branes to the $v$ direction and make them to be
parallel to a middle NS5-brane (and its mirrors also).
Note that in this case, the superpotential with finite $\mu$ has more terms as we
mentioned in the footnote \ref{more}.  
After we move the left D6-branes to the right all the way (and their
mirrors, right D6-branes to the left) past the middle NS5-brane and
the right NS5'-brane, 
the linking number $L_6$ of a single D6-brane (\ref{l6})
becomes $L_6=\frac{1}{2}-n_{4L}$, because $n_{5L}=1$ from the right NS5'-brane, 
which should be equal to the original linking number 
$L_6=-\frac{1}{2}$ because $n_{5R}=1$ from the left NS5'-brane. 
Then we should add $N_f$ D4-branes, corresponding to the meson $M_0$, to the left 
side of all the right $N_f$ D6-branes (and their mirrors) in Figure 2. 
Note that the difference, compared with previous case where $0<\theta
< \pi/2$, 
appears here when a D6-brane crosses the middle NS5-brane.
Due to the parallelness of these, there is no creation of D4-branes.

Next, we move the left NS5'-brane to the right all the way past
O6-plane
(and its mirror, right NS5'-brane to the left), and then the linking number
of NS5'-brane in (\ref{l5}), 
$L_5 =\frac{N_f}{2}-N +  N_f$ because $n_{4L}=N$, $n_{4R}=N_f$ and $n_{6L}=N_f$.
Originally, it was $L_5=-\frac{N_f}{2} +N_c$ from the fact that
$n_{4R}=N_c$ and $n_{6R}=N_f$ in Figure 1.
This implies that $N$ becomes 
\bea
N=2N_f-N_c.
\label{casetwo}
\eea
The $N_f$-dependent term in finite $\mu$ case (\ref{general}) is distributed 
as $N_f$ in previous case (\ref{caseone}) and $2N_f$ in present case (\ref{casetwo}). 
There is no D4-brane creation when NS5'-brane crosses an O6-plane
because they are parallel to each other.
Finally, we are
left with Figure 2 with $\theta=\pi/2$ 
except that the number of D4-branes connecting
between D6-branes and NS5'-brane is $N_f$ not $2N_f$ and the gauge
group is $SU(N=2N_f-N_c)$ after we rotate D6-branes to the original
positions
and make them to be parallel to NS5-brane
\footnote{For an IR free magnetic  region, we should have
  $N < N_f+2$. With this condition and $N >0$, the range for $N_f$ is given by 
$ \frac{N_c}{2}< N_f < N_c+2$.}.
These $N_f$ D4-branes correspond to the singlet $M_0$ (\ref{mesons}).
The left hand side of O6-plane including NS5-brane is exactly the same
magnetic brane configuration describing the moduli space in massive
${\cal N}=1$ SQCD, in particular, the Figure 6 of \cite{BGHSS}.
The magnetic superpotential corresponding to the electric
superpotential $W=m \tr Q \widetilde{Q}$ is given by
\bea
W_{dual} = \tr M_0 q \widetilde{s} s \widetilde{q} + \tr m M_0
\label{correctsuperpotential}
\eea 
where the first term comes from the superpotential with $M_0$ for
finite $\mu$ case in previous section (\ref{dual}) and 
the second term comes from
the mass term for the quarks in the context of magnetic theory. 
Here $q$ and $\widetilde{q}$ are fundamental and antifundamental for
the gauge group indices and antifundamentals for the flavor indices.
The fields $s$ and $\widetilde{s}$ are symmetric tensor and conjugate symmetric
tensor for the gauge group indices respectively with no flavor
indices.
Then, $q \widetilde{s} s \widetilde{q}$ has rank $N$ while $m$ has a
rank $N_f$.  Therefore, the F-term condition, the derivative the 
superpotential $W_{dual}$ with respect to $M_0$, cannot be satisfied 
if the rank $N_f$ exceeds $N$. This is so-called rank condition given 
by \cite{ISS} \footnote{Since the dual theory is in the IR free range,
the Kahler potential is regular around the origin of the field space
and can be expanded. Since the meson $M_0$ is identified with $Q
\widetilde{Q}$
in the electric description, its dimension is not equal to
1. Therefore $1/\Lambda_1^2$ appears in the Kahler potential
$M_0^{\dagger} M_0$ where $\Lambda_1$ is an electric scale. By
redefining $M_0, q, \widetilde{q}, s$, and
$\widetilde{s}$, their coefficients in the Kahler potential
are normalized to be 1. Also the coupling connecting the scale
$\Lambda_1$ and magnetic scale $\widetilde{\Lambda}_1$ and the mass
matrix $m$ are redefined appropriately. The higher order
corrections in the Kahler potential are negligible.}.    

The classical moduli space of vacua can be obtained from F-term
equations.
From the F-terms $F_{q}$ and $F_{\widetilde{s}}$, one gets
$\widetilde{s} s \widetilde{q} M_0 =0=   s \widetilde{q} M_0 q$.
Similarly, one obtains 
$\widetilde{q} M_0 q \widetilde{s} =0=   M_0 q \widetilde{s} s$ from 
the F-terms $F_{s}$ and $F_{\widetilde{q}}$. Moreover, there is a relation 
$ q \widetilde{s} s \widetilde{q} +  m=0$ from the F-term $F_{M_0}$.
From the conditions $s \widetilde{q} M_0=0= M_0 q \widetilde{s}$ which
satisfy the above four equations for nonzero vacuum expectation
values $q, \widetilde{q}, s$ and $\widetilde{s}$, one
can fix the form of $M_0$ and part of $q \widetilde{s}$ and $s
\widetilde{q}$ which can be fixed by using $F_{M_0}=0$ further and one obtains 
the following solutions 
\bea
<q \widetilde{s}> =  \left(
\begin{array}{c}
\sqrt{m} e^{\phi} {\bf 1}_{N}  \\
0
\end{array}
\right),  
<s \widetilde{q}> =
 \left(
\begin{array}{cc}
\sqrt{m} e^{-\phi}  {\bf 1}_{N}   &
0
\end{array}
\right), 
<M_0>  =
 \left(
\begin{array}{cc}
0  & 0 
 \\
0 & \Phi_0  {\bf 1}_{N_f-N} 
\end{array}
\right)
\label{vacuum}
\eea
where $\Phi_0  {\bf 1}_{N_f-N}$ is an arbitrary 
$(N_f-N) \times (N_f-N)$ matrix and the zeros of 
$< q \widetilde{s}>$ and $< s \widetilde{q}> $ are $(N_f-N) \times N $
and $N \times (N_f-N)$ zero matrices respectively. Similarly, the zeros
of $N_f \times N_f$  matrix $M_0$ are assumed also.
Then $\Phi_0$ and $(\sqrt{m} e^{\phi}, \sqrt{m} e^{-\phi})$
parametrize
a pseudo-moduli space.
Let us expand around on a point on (\ref{vacuum}) as done in \cite{ISS}. 
That is, 
\bea
q  & = &
\left(
\begin{array}{c}
q_0  {\bf 1}_{N} +\frac{1}{\sqrt{2}}(\delta \chi_{+} + \delta \chi_{-})
 {\bf 1}_{N} \nonu \\
\delta \varphi
\end{array}
\right), \quad 
\widetilde{s} = \left(\widetilde{s}_0 + \delta
  \widetilde{X} \right) {\bf 1}_{N \times N}, 
\quad
s   =   \left(s_0 + \delta
  X\right) {\bf 1}_{N \times N},
\nonu \\
\widetilde{q}  & = &
 \left(
\begin{array}{cc}
\widetilde{q}_0   {\bf 1}_{N} +
\frac{1}{\sqrt{2}}(\delta \chi_{+} - \delta \chi_{-})
  {\bf 1}_{N}   &
\delta \widetilde{\varphi}
\end{array}
\right), 
\qquad
M_0  =
 \left(
\begin{array}{cc}
\delta Y  & \delta Z
 \\
\delta \widetilde{Z} & \Phi_0  {\bf 1}_{N_f-N} 
\end{array}
\right).
\nonu
\eea
Then the superpotential becomes 
\bea
W_{dual}^{fluct} & = &  \Phi_0 \left( \delta \varphi \; \widetilde{s}_0 
\; s_0 \; \delta \widetilde{\varphi}  + m \right) +
  \delta Z \; \delta \varphi \; \widetilde{s}_0 \; s_0 \; \widetilde{q}_0
+  \delta \widetilde{Z} \; q_0 \; \widetilde{s}_0 \; s_0 \; 
\delta \widetilde{\varphi}  
\nonu \\
 & + & \left( \frac{1}{\sqrt{2}}
\delta Y \; \delta \chi_{+} \; \widetilde{s}_0 \; s_0 \; \widetilde{q}_0  
+ \cdots \right)
+ \mbox{(cubic)}
\label{s}
\eea
where $\mbox{(cubic)}$ stands for the terms that are cubic or higher
in the fluctuations.
Then to quadratic order, the model splits into two sectors where the
first piece(the first line of (\ref{s})) is an O'Raifeartaigh type model 
and the  second piece is supersymmetric and will not contribute to the
supertrace(the second line of (\ref{s})). The fields 
$\delta \chi_{\pm}, \delta Y, \delta X$ and $\delta \widetilde{X}$
couple to the supersymmetry breaking fields $\delta \varphi$ and
$\delta \widetilde{\varphi}$ via terms of cubic and higher
 order in the fluctuations.
Then the remaining relevant terms of superpotential are given by
\bea
W_{dual}^{rel} & = &  \Phi_0 \left( \delta \hat{\varphi}  
\; \delta \hat{\widetilde{\varphi}} + m \right) +
  \delta Z \; \delta \hat{\varphi} \; s_0 \; \widetilde{q}_0 
+ \delta \widetilde{Z} \; q_0 \; \widetilde{s}_0 \;
\delta \hat{\widetilde{\varphi}}
\nonu
\eea
where 
$ \delta \hat{\varphi} \equiv \delta \varphi \; \widetilde{s}_0 $ and 
$\delta \hat{\widetilde{\varphi}} \equiv s_0 \; \delta 
\widetilde{\varphi} $.
At one loop, the effective potential $V_{eff}^{(1)}$ for $\Phi_0$ 
can be obtained 
from this relevant part of superpotential which consists of 
the matrix $M$ and $N$(or $\hat{M}$ and $\hat{N}$) of \cite{Shih}. 
One can compute these matrices for simplest rank case, as in the
appendix A of \cite{Shih}. His defining function ${\cal F}(v^2)$ can be
computed
and using the relation(the equation (2.14) of \cite{Shih}) 
of $m_{\Phi_0}^2$ and ${\cal F}(v^2)$(and
noting that the seconde piece  of $m_{\Phi_0}^2$ 
will vanish), one gets 
that $m_{\Phi_0}^2$ will contain $\frac{m}{8\pi^2}(\log 4 -1) > 0$, by
taking the limit where $q_0 \widetilde{s_0} \rightarrow \sqrt{m}
e^{\phi}$
and  $s_0 \widetilde{q}_0 \rightarrow \sqrt{m}
e^{-\phi}$, as in (\ref{vacuum}).
This implies these vacua are stable.
In the brane configuration from Figure 2, since the $N$ D4-branes 
can slide along the two NS5'-branes when $\theta=\frac{\pi}{2}$, 
the vacuum expectation values
of
$s$ and $\widetilde{s}$ can be diagonalized.   

By recombination of $N$ of D4-branes connecting D6-branes and
NS5'-brane with those connecting NS5'-brane and NS5-brane and moving
them in $v$ direction (and their mirrors) from Figure 2, 
the minimal energy supersymmetry breaking brane
configuration is shown in Figure 3 \footnote{One can compute the
  energy of the nonsupersymmetric vacuum using either the effective
  field theory or the DBI action using the length of D4-branes, as
  done in \cite{BGHSS}. In the
limit of $|\Delta x| << L_0$ where $\Delta x$ is a distance of
D6-branes along the $v$ direction and $L_0$ is a distance between
D6-branes and NS5'-brane with massless flavors, the two expressions
are the same. Moreover, the energy difference between the tachyonic
state and the vacuum agrees with the same quantity from field theory
in the limit  $|\Delta x| << L_0$. } 
that was observed also in \cite{FGU}.
Some entries in the dual quarks $q, \widetilde{q}$ and dual tensors
$s, \widetilde{s}$ acquire the nonzero expectation values in terms of
the eigenvalues of mass matrix $m$ to minimize
the F-term $F_{M_0}$ in this dual gauge theory. 
When one gets the nonzero vacuum expectation values for $<M_0>, <q>$, and
$<\widetilde{q}>$, then 
the separation of the D4-branes along the middle NS5-brane  
corresponds to the mass of two index tensor $(s \widetilde{s})$.
On the other hand, for the nonzero expectation values of 
$<M_0>, <s>$ and $<\widetilde{s}>$, 
the separation of the D4-branes along the middle NS5-brane  
corresponds to the mass for the dual quarks $(q \widetilde{q})$.

\begin{figure}[ht]
   \epsfxsize=6in 
\centerline{\epsffile{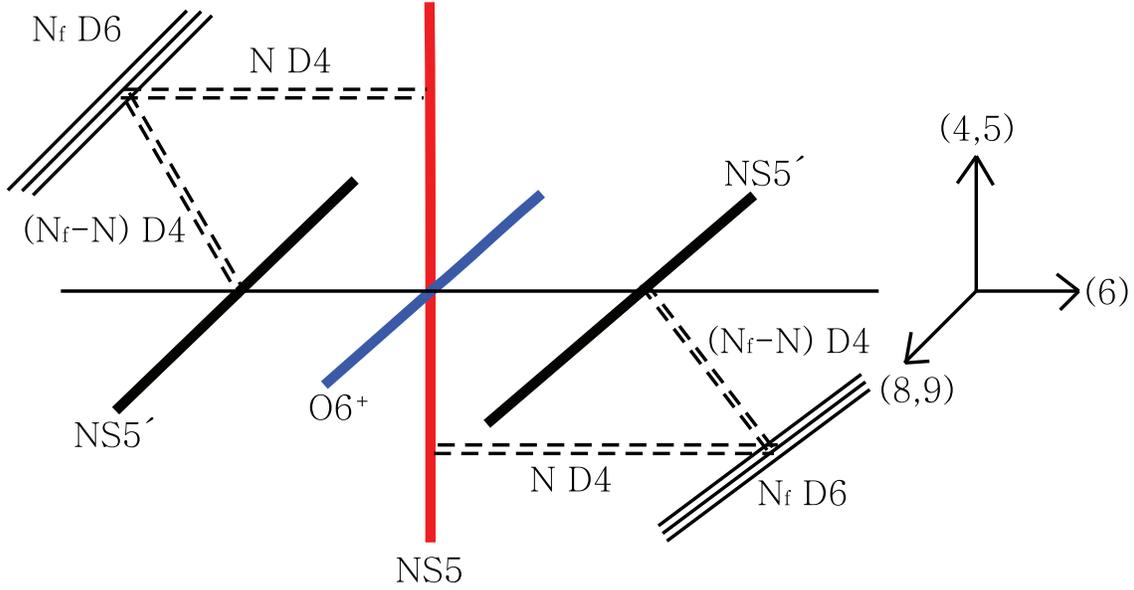}}
   \caption[FIG. \arabic{figure}.]{ 
The nonsupersymmetric minimal energy
brane configuration for the $SU(N=2N_f-N_c)$ gauge theory  with 
a symmetric flavor $s$, a conjugate symmetric flavor $\widetilde{s}$ and 
$N_f$ fundamental flavors $q, \widetilde{q}$.
 The $N_f$ D4-branes connecting between
NS5'-brane and D6-branes are the dual gauge singlet
corresponding to the meson $M_0$ (\ref{mesons}).
When we consider the M-theory lift of this brane configuration,
we move the left NS5'-brane  to the
$v$ direction holding everything else fixed, instead of moving D6-branes.
The corresponding mirrors and D4-branes are placed appropriately
during this process.}
\label{fig3}
\end{figure}

\section{M-theory description of 
nonsupersymmetric meta-stable brane configuration}

The M5-brane spans (0123) directions and wraps on a Riemann surface 
inside (4568910) directions. 
The Taub-NUT space (45610) is parametrized by 
two complex variables $(v,y)$ and flat two-dimensions (89) are by
a complex variable $w$.
The mass dimensions of these variables are
given by 
$
[v]=1, [y]=2N_c,  [w]=2
$
respectively \cite{BGHSS}. The mass dimension for $v$ can be seen from the
corresponding Seiberg-Witten curve.
For large $v$, since $w$ goes to $ \pm \mu v$ for the left and right
$NS5_{\pm \theta}$-branes where 
$\mu$ is a mass of the adjoint chiral multiplet, the mass dimension of
$w$ is equal to 2.
The mass dimension of $y$ 
can be determined also by 
the boundary condition near 
$w =\infty$. 

The precise relation between the holomorphic coordinates $(v,y,w)$ and 
physical coordinates (4568910)
are given by
\bea
v =\frac{x^4 + i x^5}{\ell_s^2}, \quad 
y =
\mu^{2N_c} e^{\frac{x^6-L_0+i x^{10}}{2R}} 
\left(\frac{\sqrt{(x^4)^2+(x^5)^2+(x^6)^2}+x^6}{R}\right)^{\frac{N_f}{2}}, \quad
w= \frac{x^8 + i x^9}{R \ell_s^2}
\nonu
\eea
where $R$ is a radius of eleventh direction and is given in terms of a
string coupling and a string scale by $R=g_s \ell_s$ and we put $R$
dependence here explicitly for M-theory description. As $R$ goes to
zero, one gets the type IIA brane description in previous section.
Note that the presence of $\mu$ term in $y$ provides the correct mass
dimension above.

One of the complex structures of Taub-NUT
space can be described by embedding it in a three complex dimensional
space
with coordinates $(x,t,v)$. 
The M5-brane curve corresponding to the type IIA brane configuration 
shown in Figure 1, in a background space \cite{Witten} of
\bea
x t = (-1)^{N_f} v^4 \prod_{k=1}^{N_f} (v^2 -e_k^2)
\label{xt}
\eea
where $e_k$ is the position of the D6-branes in the $v$ direction and
the O6-plane ${\bf Z}_2$ symmetry action we discussed before can be rewritten as 
$(x,t,v) \rightarrow (t,x,-v)$ \footnote{In a complex 3-dimensional
  space ${\bf C}^3$, the equation is characterized by $x t =
  \prod_{k=1}^{2N_f+4} (v-m_k)$ with $m_k=-e_k$. Now we are plugging the positions for
  the D6-branes including the contribution from the O6-plane. Then this can be written in
  terms of (\ref{xt}). Note that the position of O6-plane is located at $v=0$.  },
can be characterized by \cite{LLL}
\bea
t^3 + t^2 v^{N_c} + t v^2 (-1)^{N_c} v^{N_c} \prod_{k=1}^{N_f} (v-e_k)
+ v^6 \left[ \prod_{k=1}^{N_f} (v-e_k) \right]^2 (-1)^{N_f}
\prod_{k=1}^{N_f} (v+e_k) =0.
\label{cubic}
\eea
Since the location of D4-branes in $v$ direction is given by $v=0$,
the $N_c$-dependent terms in $t^2$ and $t$ above have a simple form. 
For fixed $x$, the coordinate $t$ corresponds to $y$ and for fixed
$y$, the coordinate $x$
corresponds to $1/t$. Since the O6-plane plays the role of 4
D6-branes and D6-branes lift to the Taub-NUT space in the context of
M-theory description, 
the power of $v$ in the right hand side above (\ref{xt}) 
is equal to 4
\footnote{Of course, if we consider the negative charge for the
  O6-plane, this power will be $-4$. }.
The above cubic equation (\ref{cubic}) is a polynomial of degree 3 in $t$ and this
implies that there exist 3 solutions for $t$ corresponding to three
NS5-branes we are considering \footnote{As observed in \cite{LLL}, the
curve can be factorized under the particular circumstance for the
coefficient functions appearing in the cubic equation:divisibility by
$(t+t_0)$ where $t_0$ is $v$-independent. 
In this
   case, the quadratic piece in $t$ corresponds to the Seiberg-Witten 
curve of ${\cal N}=2$ $SO(N_c)$ gauge
theory with $N_f$ flavors while the linear piece in $t$ corresponds to the
factorized middle NS5-brane. For the negative O6-plane charge, the
gauge group will be changed into $Sp(N_c)$ with $N_f$ flavors. The
M-theory descriptions in the presence of 
orientifold 4-plane \cite{Ahn97,Ahn97-1} are reproduced from the
result of \cite{AOT}. In this sense, the asymptotic behavior of
M5-brane curve for the present gauge theory 
looks similar to those of symplectic or orthogonal gauge
groups.}.

At $g_s \neq 0$, the NS5-branes can bend due to their interactions
with the D4-branes and D6-branes.
Let us consider the asymptotic behavior such that the rotated curve 
should have at $v \rightarrow \infty$ and $v=0$.
One can read off the behaviors of the left, middle, right NS5-branes
respectively
by considering the first two terms, the second and third terms, the
last two terms from above cubic equation (\ref{cubic}) respectively.
Then the behavior of the supersymmetric M5-brane curves \footnote{Without
O6-plane, the analysis for the two factor gauge groups \cite{GP} leads
to the fact that the M5-brane curve \cite{HOO} for 
${\cal N}=1$ SQCD with massive equal flavor masses
can be obtained from the special limit of vanishing gauge coupling
constant for the right factor, as expected. From the general result
of \cite{GP}, in particular, (4.39) and (4.40) of \cite{GP}, 
the appropriate orientifold projection provides the
corresponding the explicit relations of $v=v(w)$ and $y=y(w)$ for our
gauge theory.       } can be
summarized as follows:

1. $v \rightarrow \infty$ limit implies
\bea
w \rightarrow 0, \quad y \rightarrow    -(-1)^{N_c} 
\Lambda_{N=1}^{2N_c-N_f-2} v^{N_f+2} + \cdots \quad
\mbox{NS asympt. region}.   
\nonu
\eea

2.  $w \rightarrow \infty$ limit implies
\bea
v & \rightarrow &   m_f, \quad 
y \rightarrow -(-1)^{N_f-N_c}
\Lambda_{N=1}^{4N_c-4N_f-8} w^{2N_f-N_c+4}
 +\cdots
\quad \mbox{$NS_{L}'$ asympt. region}, 
\label{class2one} \\
v & \rightarrow &  -m_f, \quad  
y \rightarrow -w^{N_c}
+\cdots
\quad \mbox{$NS_{R}'$ asympt. region}. 
\label{class2two}
\eea
Here we inserted the appropriate scale $\Lambda_{N=1}$ in order to
match the mass dimension above. 
Along the line of \cite{BGHSS,Ahn06-1}, the left NS5'-brane is moved to the
$v$ direction holding everything else fixed, 
instead of moving D6-branes (and their mirrors).
In this process, the corresponding mirrors are placed in appropriate way. 
We also shift the origin for $x^6$ direction and put the left
D6-branes to the origin and denote the distance between the left
D6-branes and NS5-brane as $L_0+\Delta L$. That is, we denote the
distance between the left D6-branes and NS5'-brane by $L_0$ and the distance 
between NS5'-brane and NS5-brane by $\Delta L$, as in \cite{BGHSS}.  

3. The map between the holomorphic and physical coordinates
requires 
the condition \cite{BGHSS}
\bea
 y=0 \qquad \mbox{only if} \qquad v=0.
\label{class3}
\eea

Since the nonsupersymmetric brane configuration in Figure 3 implies that 
only the NS5'-branes are deformed by turning on the mass for the
quarks, only NS5'-branes are nonholomorphic.
The remaining NS5-brane and D6-branes remain unchanged.
The ansatz for this nonholomorphic 
curve corresponding to two NS5'-branes where they have different
boundary conditions characterized by (\ref{class2one}) and
(\ref{class2two}) 
respectively can be made as follows \cite{BGHSS}:
\bea
x^4 = f(s), \qquad x^5 =0, \qquad 
x^8 + i x^9 = e^{i \frac{x^{10}}{2N_c R}} g(s), \qquad
x^6 =s.
\nonu
\eea
The unknown functions $f(s)$ and $g(s)$ 
can be determined by solving 
Euler-Lagrange equations  for the action of surface
parametrized by $x^6$ and $x^{10}$ directions.
The metric by the six dimensional transverse directions is given by 
the Taub-NUT space (45610)  and flat two-dimensions (89) \footnote{
The metric by these six-dimensional transverse directions 
is given by 
$
ds_6^2 = G_{AB} dX^A dX^B = V d \vec{r}^2 + V^{-1}
\left( d x^{10} + \vec{\omega} \cdot d \vec{r} \right)^2
+ (dx^8)^2 + (dx^9)^2 $
where (456) directions are parametrized by $\vec{r}$ and the relation
between the harmonic function
$V$ sourced by the D6-branes  
and the vector potential 
$\vec{\omega}$ is  given by  
$\nabla \times \vec{\omega} =\nabla V$, as usual.
}.
The harmonic function appearing in the Taub-NUT space, sourced by the the left
coincident $N_f$ D6-branes located at $x^6=0$, O6-plane (which is equivalent to 4
D6-branes) 
located at
$x^6=\Delta L +L_0$, and the right coincident $N_f$ D6-branes located at
$x^6=2(\Delta L +L_0)$,
can be written as explicitly, by putting the right charges and the
locations of them,
\bea
V(s) = 1  + 
\frac{N_f R}{\sqrt{f(s)^2 + s^2}}+ 
\frac{4 R}{\sqrt{f(s)^2 + (s-\Delta L -L_0)^2}}+
\frac{N_f R}{\sqrt{f(s)^2 + (s-2\Delta L -2L_0)^2}}.
\label{harmonic}
\eea
Note that when we compare with the usual ${\cal N}=1$ SQCD with
massive flavors developed in \cite{BGHSS}, the last two terms are an
extra piece coming from the effect of O6-plane and mirrors of
D6-branes in our gauge theory.   

When the left NS5-brane is located at $v =\frac{\Delta x}{\ell_s^2}$
from Figure 3 and
$
f_L(s) =  \Delta x
$
which satisfies $f_L''(s)=0$ because $f_L(s)$ doesn't depend on $s$, 
then the equation (A.3) of \cite{BGHSS} which is valid for arbitrary
form for $V(s)$
implies that $g_L'(s) =\frac{V}{2N_c R} g_L(s)$ with
$V(s)=V(s)|_{f_L(s)=\Delta x}$.
Then, this first order differential equation leads to the following
solution for $g_{L}(s)$
\bea
g_L(s)  & = &   R \ell_s^2 \left[ (-1)^{-N_f+N_c-1} \Lambda_{N=1}^{-4N_c+4N_f+8} 
\mu^{2N_c} e^{\frac{-L_0}{2R}} \left(\frac{1}{R} \right)^{\frac{N_f}{2}} 
\right]^{\frac{1}{2N_f-N_c+4}} 
\nonu \\
&& \times 
\left(s-l + \sqrt{(\Delta
    x)^2 
+ (s-l)^2} \right)^{\frac{2}{N_c}}
 \prod_{j=0}^{1} \left[ s-2lj + \sqrt{(\Delta
    x)^2 
+ (s-2lj)^2} \right]^{\frac{N_f}{2N_c}}
\nonu
\eea
with $l \equiv \Delta L +L_0$. The $s$-independent integration
    constant is fixed by the boundary condition 
$y \rightarrow -(-1)^{N_f-N_c}
\Lambda_{N=1}^{4N_c-4N_f-8} w^{2N_f-N_c+4}
 +\cdots$ given by the classification 2 above (\ref{class2one}). 
This is a simple solution $v =\frac{\Delta
    x}{\ell_s^2}=m_f$ and $y = -(-1)^{N_f-N_c}
\Lambda_{N=1}^{4N_c-4N_f-8} w^{2N_f-N_c+4}$.
Even if $\Delta x$ is equal to zero, the function $g_L(s)$ does not vanish.
This implies $w$ does not vanish and therefore $y$ is not equal to
zero. So this is, in fact, a contradiction with the above
    classification 3 in (\ref{class3}). 
In other words, the extra piece in the potential doesn't eliminate
the instability  from a new M5-brane mode
occurring  at some point during the continuation to
M-theory description of SQCD.

Furthermore, if 
the right NS5-brane is located at $v =-\frac{\Delta x}{\ell_s^2}$ and
$
f_R(s) =  -\Delta x$ (note that this right NS5-brane is a mirror of
the left NS5-brane) from Figure 3,
then it is easy to see that
\bea
g_R(s)  & = &  (-1)^{-\frac{1}{N_c}} R^{1-\frac{N_f}{2N_c}} 
\ell_s^2 \mu^2 e^{\frac{s-L_0}{2N_cR}} 
\nonu \\
&& \times \left(s-l + \sqrt{(\Delta
    x)^2 
+ (s-l)^2} \right)^{\frac{2}{N_c}}
 \prod_{j=0}^{1} \left[ s-2lj + \sqrt{(\Delta
    x)^2 
+ (s-2lj)^2} \right]^{\frac{N_f}{2N_c}}
\nonu
\eea
where $l$ was defined before. 
In this case, we use the second property of the classification 2 given
in
(\ref{class2two}) 
for the boundary condition. Since the potential form of $V(s)$ is
given by the same expression, we are left with the same contradiction
we met before.

Instead of imposing the boundary condition 
at large $s$, we require that M5-branes end on the D6-branes: 
$f(s)$ at $s=0$ vanishes.
For the case of
$
f_L(s) = c s
$
which still satisfies $f_L''(s)=0$ and $f_L'(s) \neq 0$,
the (A.3) of \cite{BGHSS} leads to the fact that there exists 
$
g_L'(s) = \frac{\sqrt{1+f_L'(s)^2}}{2N_c R} V g_L(s)
$
and when $L=L_0$ and $c=\frac{ \Delta x}{L_0}$, 
this straight line solution 
will become the type IIA brane configuration  
in the $R \rightarrow 0$ 
limit. 
However, the behavior at infinity is different from the above 
classification 1 and 2. Similar analysis for the $f_R(s)$ where a
straight line has an expression $f_R(s)=c s + d$ can be done.
Therefore, the supersymmetric brane configuration and the nonsupersymmetric 
brane configuration are vacua of different theories because
the boundary conditions at infinity are different.

In \cite{BGHSS}, they have tried to search for the possibility for the
other solutions with the right boundary conditions by substituting the
explicit form for the potential and have obtained third order nonlinear
differential equation (A.5) of
\cite{BGHSS}. Surprisingly, the exact solution for the $f(s)$ with
three parameters was
found through (A.5)-(A.10) of \cite{BGHSS}. However, since our potential has
an extra piece in (\ref{harmonic}), compared with the one in
\cite{BGHSS}, 
it is evident that the third order nonlinear
differential equation for $f(s)$ cannot be solved exactly. This
feature have occurred also in an example of \cite{Ahn06-1}. The above
solutions $f_{L,R}(s) =\pm \Delta x$, $f_{L}(s) = c s$ and $f_R(s)= c
s +d$ are 
particular subfamily of the general solutions
and, in principle, there could exist a solution 
having the correct boundary conditions both at infinity 
and at the D6-brane with $f_{L,R}''(s) \neq 0$. It seems to be difficult to
construct this solution even if one uses the numerical analysis for
the complicated diffferential equation corresponding to (A.5) of 
\cite{BGHSS}, as in \cite{Ahn06-1}.

\section{Conclusions and outlook}

We have constructed the type IIA brane configuration, presented in
Figure 3,
corresponding to 
the meta-stable supersymmetry breaking vacua for 
${\cal N}=1$ $SU(N_c)$ supersymmetric gauge
theory with a symmetric flavor, a conjugate symmetric flavor and
fundamental flavors. In doing this, the appropriate brane motion from
the electric configuration to magnetic configuration has played the
crucial role  in order to obtain the right superpotential 
(\ref{correctsuperpotential}) that gives rise to the breaking of the supersymmetry. 
The O6-plane doesn't contribute to the creation of D4-branes because
it is parallel to NS5'-brane so that 
the rank for the dual magnetic gauge group doesn't contain the
constant term, contrary to the finite mass of adjoint field.   

It is natural to ask whether the method for infinity limit of adjoint mass 
can apply to other supersymmetric gauge theories which can be realized
in terms of type IIA string theory.
As already observed in \cite{FGU}, for example, it is an open problem
to construct the nonsupersymmetric brane configuration corresponding
to
the meta-stable supersymmetry breaking vacua for 
${\cal N}=1$ $SU(N_c)$ gauge theory with a conjugate symmetric flavor, an
antisymmetric flavor, fundamental flavors, and anti-fundamental
flavors. 
This theory has a set of flat directions and the gauge group is
broken to either $SO(n)$ with an antisymmetric (adjoint) tensor and vectors  
or  $Sp(n)$ with a symmetric (adjoint) tensor and fundamentals \cite{ILS}.
We expect the meta-stable nonsupersymmetric vacua for
these orthogonal or symplectic  gauge theories
should lead to the mesonic deformations \cite{AGM,AGM1} of the ${\cal N}=1$
$SU(N_c)$ SQCD with an adjoint field,  
by adding the appropriate orientifold 4-plane.  

There exist many SQCD-like theories \cite{ILS} where the
Seiberg dual theories are known explicitly. It would be interesting to
find out how to 
extract the simplified superpotential like as
(\ref{correctsuperpotential}) by looking at its brane motion
intuitively without dealing with various meson
fields directly from gauge theory side.

\vspace{.7cm}

\centerline{\bf Acknowledgments}

We would like to A. Hanany, K. Landsteiner and D. Shih for discussions.
This work was supported by grant No.
R01-2006-000-10965-0 from the Basic Research Program of the Korea
Science \& Engineering Foundation.

\end{document}